\documentclass[12pt,a4paper,dvips]{article}

\usepackage{graphicx}
\usepackage{amsbsy}
\usepackage{amssymb}

\begin{document}

\newcommand{\be}{\begin{equation}}
\newcommand{\ee}{\end{equation}}
\newcommand{\bea}{\begin{eqnarray}}
\newcommand{\eea}{\end{eqnarray}}
\newcommand{\tr}{\,\hbox{tr }}
\newcommand{\Tr}{\,\hbox{Tr }}
\newcommand{\Det}{\,\hbox{Det }}
\newcommand{\fslash}{\hskip-.2cm /}
\title{An Improved Gaussian Approximation for Quantum Field Theory}

\author{
Antun Bala\v z
\thanks{http://www.phy.bg.ac.yu/ctp/pft/a\_balaz/}\\
Aleksandar Beli\'c
\thanks{http://www.phy.bg.ac.yu/ctp/cmt/a\_belic/}\\
and\\
Aleksandar Bogojevi\'c\vspace{0.25cm}
\thanks{http://www.phy.bg.ac.yu/ctp/pft/a\_bogojevic/}\\
\textsl{Institute of Physics}\\
\textsl{P.O.B. 57, Belgrade 11001}\\
\textsl{Yugoslavia}\\
}

\date{Preprint IP-HET-98/11\\
April 1998}
\maketitle

\begin{abstract}
We present a new approximation technique for quantum field theory.
The standard one-loop result is used as a seed for a recursive formula
that gives a sequence of improved Gaussian approximations for the
generating functional. In a different setting, the basic idea of this
recursive scheme is used in the second part of the paper to substantialy
speed up the standard Monte Carlo algorithm.
\end{abstract}
\newpage

\section{Introduction}

Quantum field theory is compactly written in terms of path integrals. Path
integrals have been specially usefull in dealing with symmetries, quantizing
gauge theories, etc. At some point, however, we need to calculate the path
integrals. The problem is that the only path integrals that we know how to
solve correspond to free theories (Gaussian integrals). We would be in a very
sorry state if we didn't have a generic approximation scheme at our disposal.
Semi-classical or loop expansion is just such an approximation scheme. In
fact, much of what we know about quantum field theory comes from one-loop
results. The one-loop result is obtained by Taylor expanding the action
around classical fields and disregarding cubic and higher terms. In this
way the path integral is approximated by a Gaussian.

In this paper we will consider another Gaussian approximation to the path
integral. Unlike the one-loop result, here we will Taylor expand about the
average field $\varphi\equiv\langle\phi\rangle$. We shall show that this
leads to an improved approximation given in terms of a recursive relation.

\section{The Gaussian Approximation}

The central object in quantum field theory is the generating
functional $Z[J]$. Functional derivatives of $Z[J]$ with respect to the
external fields $J(x)$ give the Green's functions of the theory. The
generating functional is determined from the (Euclidian) action $S[\phi]$
through the path integral
\be
Z[J]=\int [d\phi]\,e^{-\frac{1}{\hbar}
\left(S[\phi]-\int dx\,J(x)\,\phi(x)\right)}\ .
\ee
The integration measure is, formaly, simply
\be 
[d\phi]=\prod_{x\in\mathbb{R}^d}d\phi(x)\ ,
\ee
where $d$ is the dimension of space-time. We are interested in looking at a
set of approximations to the above path integral. The approximations are
valid in all $d$. In this letter, however, we are going to look at the
simpler case of $d=0$ theories, where it is easy to compare our results
with exact numerical calculations. In $d=0$ functionals become functions,
and the path integral reverts to a single definite integral over the whole
real line.
\be
Z(J)=\int d\phi\,e^{-\frac{1}{\hbar}
\left(S(\phi)-J\,\phi\right)}\ .\label{z}
\ee
An even more useful object is $W(J)$ --- the generator of connected diagrams,
defined by
\be
Z(J)=Z(0)\,e^{-\frac{1}{\hbar}W(J)}\ .
\ee
In statistical mechanics parlance this is the free energy. The quantum
average of the field $\phi$ is
\be
\varphi\equiv\langle\phi\rangle=-\,\frac{\partial}{\partial J}\,W(J)
\ .\label{average}
\ee
In the Gaussian approximation, we Taylor expand the action in the path
integral around some reference point $\phi_\mathrm{ref}$, and keep terms that are at
most quadratic in $\phi-\phi_\mathrm{ref}$. Thus, we use
\be
S(\phi)\approx S(\phi_\mathrm{ref})+S'(\phi_\mathrm{ref})\,
(\phi-\phi_\mathrm{ref})+
\frac{1}{2}\,S''(\phi_\mathrm{ref})\,(\phi-\phi_\mathrm{ref})^2\ .
\ee 
The integral in (\ref{z}) is now a Gaussian and we find, up to an unimportant
constant, that
\be
W_\mathrm{Gauss}(J,\phi_\mathrm{ref})=S(\phi_\mathrm{ref})-
J\,\phi_\mathrm{ref}+
\frac{\hbar}{2}\,\ln S''(\phi_\mathrm{ref})-
\frac{1}{2}\,\frac{(S'(\phi_\mathrm{ref})-J)^2}{S''(\phi_\mathrm{ref})}\ .
\label{w-g}
\ee
For this approximation to make sense, the integral must get its dominant
contribution from the vicinity of the reference point $\phi_\mathrm{ref}$.
The standard Gaussian approximation corresponds to the choice
$\phi_\mathrm{ref}=\phi_\mathrm{class}\,(J)$, where
$\phi_\mathrm{class}$ is the solution of the classical equation of motion
$S'=J$. The classical solution is the maximum of the integrand in
(\ref{z}). This specific choice of $\phi_\mathrm{ref}$ gives us the
standard one-loop result
\be
W(J)\approx W_1(J)\equiv S(\phi_\mathrm{class})-J\,\phi_\mathrm{class}+
\frac{\hbar}{2}\,\ln S''(\phi_\mathrm{class})\ .
\ee
As is well known, loop expansion is just an expansion in powers of $\hbar$.
The one-loop result gives us the first quantum correction to classical
physics. From now on we set $\hbar=1$.

\section{Improving the Gaussian Approximation}

In this section we will choose a \emph{different} expansion point
$\phi_\mathrm{ref}$ for our general Gaussian formula (\ref{w-g}).
The idea is to expand around the average field $\varphi$.
Although the classical solution gives
the maximum of the integrand, expansion around $\varphi$ gives a better
approximation for the area under the curve. This is in particular true for
large values of $J$. We will work with $\phi^4$ theory in $d=0$, whose
action is given by
\be
S(\phi)=\frac{1}{2}\,\phi^2+\frac{1}{4!}\,g\,\phi^4\ .\label{phi4}
\ee
The classical equation of motion is now a cubic algebraic
equation. We easily find the unique real solution. In this way we get a
closed form expression for the one-loop approximation $W_1(J)$. The
Gaussian approximation around the average field $\varphi$ is simply
\be
S(\varphi)-J\,\varphi+
\frac{1}{2}\,\ln S''(\varphi)-
\frac{1}{2}\,\frac{(S'(\varphi)-J)^2}{S''(\varphi)}
\ .\label{w_q}
\ee
To be able to calculate this \emph{in closed form} we need to know
$\varphi(J)$, which is tantamount to knowing how to do the theory exactly,
since $\varphi$ and its derivatives give all the connected Green's functions.
The use of equation (\ref{w_q}) comes about when one solves it iteratively.
We use equations (\ref{average}) and (\ref{w-g}) as the basis for the
following iterative process
\be
\varphi_{n+1}(J)=-\,\frac{d}{dJ}\,W_\mathrm{Gauss}(J,\varphi_n(J))\ .
\label{phi-n}
\ee
Differentiating (\ref{w-g}) we find
\bea
\lefteqn{
\varphi_{n+1}=
\varphi_n-\,\frac{S'(\varphi_n)-J}{S''(\varphi_n)}\,-}\nonumber\\
& &{}\quad\quad-\,\frac{S'''(\varphi_n)}{2}\,
\left(\left(\frac{S'(\varphi_n)-J}
{S''(\varphi_n)}\right)^2+
\,\frac{1}{S''(\varphi_n)}\right)\frac{d\varphi_n}{dJ}\ .
\label{recurse}
\eea
For the seed of this iteration we choose the classical field, i.e.
$\varphi_0=\phi_\mathrm{class}$. In this way we obtain a sequence of points
$\varphi_0,\varphi_1,\varphi_2,\ldots$ or equivallently of approximations
to the connected generating functional
$W_1,W_2,W_3,\ldots$ given by
$W_{n+1}(J)=W_\mathrm{Gauss}(J,\varphi_n(J))$. The idea behind this is
obvious --- we want to obtain a sequence of $W_n(J)$'s that
give better and better approximations to $W(J)$. The following two figures
show that this realy works.

\begin{figure}[!ht]  
    \centering
    \includegraphics[height=5.5cm]{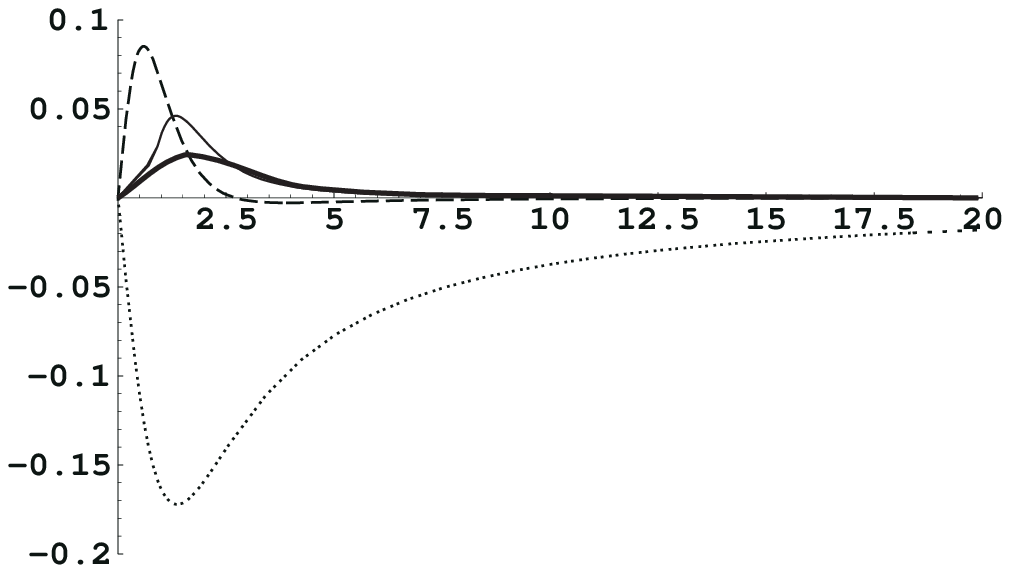}
    \caption{Plots of $\varphi-\varphi_0$ (dotted line),
    $\varphi-\varphi_1$ (dashed line), $\varphi-\varphi_2$ (thin line)
    and $\varphi-\varphi_\infty$ (thick line) as functions of $J$.
    Here we had $g=1$.}
    \label{4k}
\end{figure}

\begin{figure}[!ht]
    \centering
    \includegraphics[height=5.5cm]{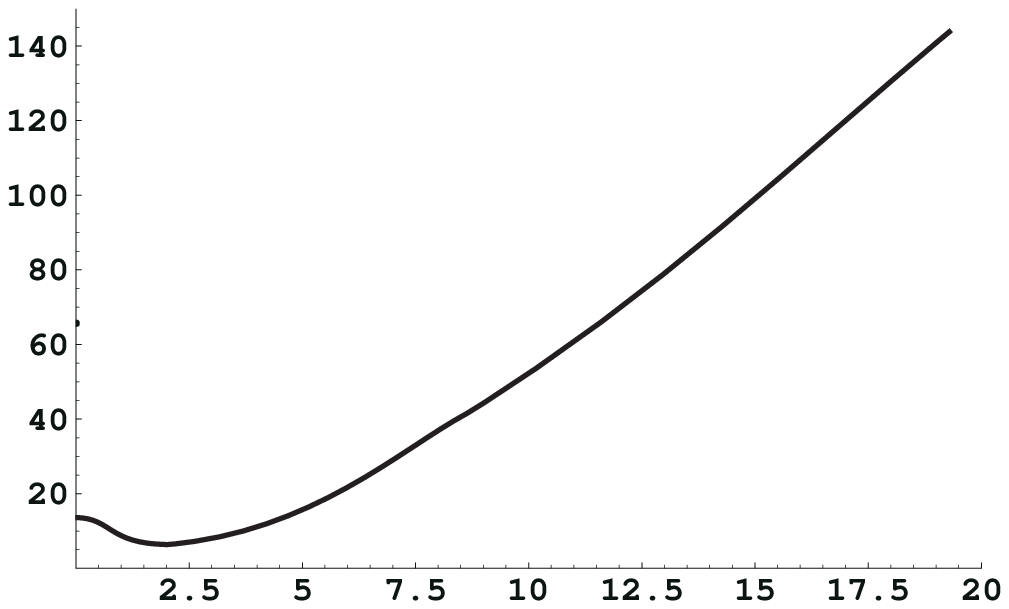}
    \caption{Plot of 
    the ratio $|(\varphi-\varphi_0)\,/\,(\varphi-\varphi_\infty)|$ as a
    function of $J$. This is a direct measure of how the new
    approximation outperforms the standard one-loop result.
    Here we had $g=1$.}
    \label{relinv} 
\end{figure}

From Figure~\ref{4k} we see that the sequence of $\varphi_n$'s converges
to $\varphi_\infty$
\footnote{In iterating (\ref{recurse}) we necessarily discretize the $J$'s.
The coursness of this discretization effects the speed of convergence of
the $\varphi_n$'s. The standard way out of this problem is to introduce a
small mixing papameter $\epsilon$. Instead of the recursive relation
$\varphi_{n+1}=f(\varphi_n)$ one then consideres
$\varphi_{n+1}=(1-\epsilon)\varphi_n+\epsilon f(\varphi_n)$.}.
Note that $\varphi_\infty\ne \varphi$. The reason for this is obvious: We
used the Gaussian approximation $W_\mathrm{Gauss}$ in defining our recursive
relation, and there is no reason to expect that this converges to the exact
result. It does, however, converge and $\varphi_\infty$ represents an
excellent approximation to $\varphi$. To see this, in Figure~\ref{relinv}
we have ploted the ratio
$|\frac{\varphi-\varphi_0}{\varphi-\varphi_\infty}|$. This ratio represents
a direct measure of the improvement of approximations in going from the
one-loop result $W_1(J)=W_\mathrm{Gauss}(J,\varphi_0(J))$ to our improved
Gaussian result $W_\infty(J) =W_\mathrm{Gauss}(J,\varphi_\infty (J))$.

As we have already mentioned, our new approximation was tailored to work
well for large $J$. This can be read off directly from Figure~\ref{relinv}.
For example, for $J=15$ the improved Gaussian is about one hundred times
better than the one-loop result. The new approximation is poorest for
$J\approx 2$, but even there it beats the old approximation by a factor
of seven. Most of the time we are interested in working with small or
zero external fields. In the vicinity of $J=0$ the new approximation
is fourteen times better than the old one.

\section{Monte Carlo}

The aim of our investigations so far has been to develop better analytic
approximation schemes that can be applied to general quantum field theories.
We worked in $d=0$ in order to be able to make a simple comparison with
exact (numerical) results. In this section we will look at the numerical
techniques themselves. We used the Monte Carlo algorithm \cite{kw} for
calculating path integrals. In $d=0$ Monte Carlo is not the most efficient
way to do things --- its advantages become apparent as we look at larger
and larger numbers of integrations. We use Monte Carlo in order to
investigate the algorithm itself in light of what we have learned in
the previous two sections.

We start with a brief introduction of the method. In order to calculate
the definite integral $\int f(\phi)d\phi$ we choose a non-negative function
$p(\phi)$ normalized so that $\int p(\phi)\,d\phi=1$. Therefore, $p(\phi)$
is a probability distribution. The integral is now
\be
\int f(\phi)\,d\phi=\int \frac{f(\phi)}{p(\phi)}\,p(\phi)\,d\phi\equiv
\left\langle\frac{f}{p}\right\rangle _p\ ,
\ee
where $\langle F\rangle_p$ represents the mean value of $F$ with respect
to the probability distribution $p$. Therefore, the integral of $f$ is
given as the mean value of $f/p$ on a sample of random numbers whose
probability distribution is given by $p$. In practice, this mean value
is estimated using a finite number $N_\mathrm{mc}$ of Monte Carlo samples,
and the error of such an estimate is itself estimated to be
$\sigma_{f/p}=\sqrt{\sigma_{f/p}^2}\,$, where the variance equals
\be
\sigma^2_{f/p}=\frac{\left\langle
\left(\frac{f}{p}\right)^2\right\rangle_p-
\left\langle \frac{f}{p}\right\rangle^2_p}{N_\mathrm{mc}-1}\ .
\ee
The central limit theorem guarantees that the Monte Carlo algorithm
converges to $\int f\,d\phi$ for an arbitrary choice of distribution
$p$. The only condition that must be met is $\sigma^{2}_{f/p}<\infty$.
This freedom in the choice of $p$ is used to speed-up the convergence of
the algorithm. The speed of convergence is measured by the efficiency
$\cal E$, given by
\be
{\cal E}=\frac{1}{T\,\sigma^{2}_{f/p}}\ ,
\ee
where $T$ represents the total computation time. Note that a hundred
fold increase of efficiency corresponds to one extra significant figure
in the final result.

In our calculation we chose $p(\phi)$ to be the Gaussian normal distribution
\be
p(\phi)=\frac {1}{\sqrt{2\pi\sigma^2}}\,
\exp\left(-\,\frac{(\phi-a)^2}{2\sigma^2}\right)\ ,
\ee
where $a$ and $\sigma$ completely determine the distribution. There are two
reasons for using this distribution. First of all, the function we are
integrating can be approximated by a Gaussian over a wide range of
parameters $J$ and $g$. A good choice of $a$ and $\sigma$ makes $f/p$
almost constant over the range of integration, thus making the variance
small. Second, there exists a specific algorithm for generating random
numbers conforming to a Gaussian distribution. The Box-Muller algorithm
\cite{ptwf} is much more efficient than the standard Metropolis algorithm
\cite{mrrtt} since it doesn't give rise to autocorrelations of generated
numbers. In the Metropolis algorithm autocorrelations can be pronounced,
and their removal substantialy slows down the simulation.

The choice of probability distribution has a great effect on the efficiency.
For example, the efficiency corresponding to the uniform distribution on
the interval $\phi\in [-100,100]$ is 3.5 $10^{10}$ times smaller than the
efficiency achieved by the Gaussian distribution centered at
$a=\phi_\mathrm{class}$ with optimal choice of width $\sigma$. Having chosen
$p$ to be a Gaussian, the computation time $T$ depends only on the number of
Monte Carlo samples $N_\mathrm{mc}$. Therefore, in our case, maximalization
of efficiency is equivallent to a minimalization of the variance $\sigma_{f/p}^2$.

\begin{figure}[!ht]
  \centering
  \includegraphics[height=5.5cm]{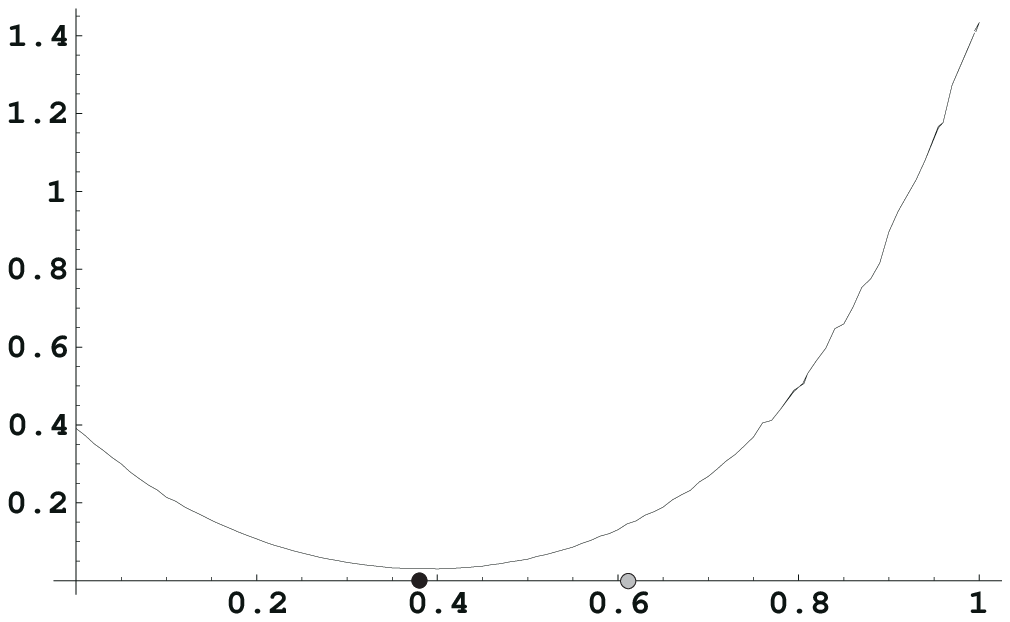}
  \caption{The variance as a function $a$. The plot is for $g=10$, $J=1$.
  The variance is minimized for $\varphi(1)=0.376799$ (black dot).
  The classical field is $\phi_\mathrm{class}(1)=0.614072$ (grey dot).}
  \label{min}
\end{figure}

In the previous sections we saw that it is even better to expand around
$\varphi$. In the Monte Carlo setting this should translate into a further
increase in efficiency. This is \emph{precisely} what we see. By varying
the center of the Gaussian $a$ (always using optimal width for
that given $a$), we find maximum efficiency precisely at $a=\varphi$ as we
can see in Figure~\ref{min}.

\begin{figure}[!ht]
  \centering
  \includegraphics[height=5.5cm]{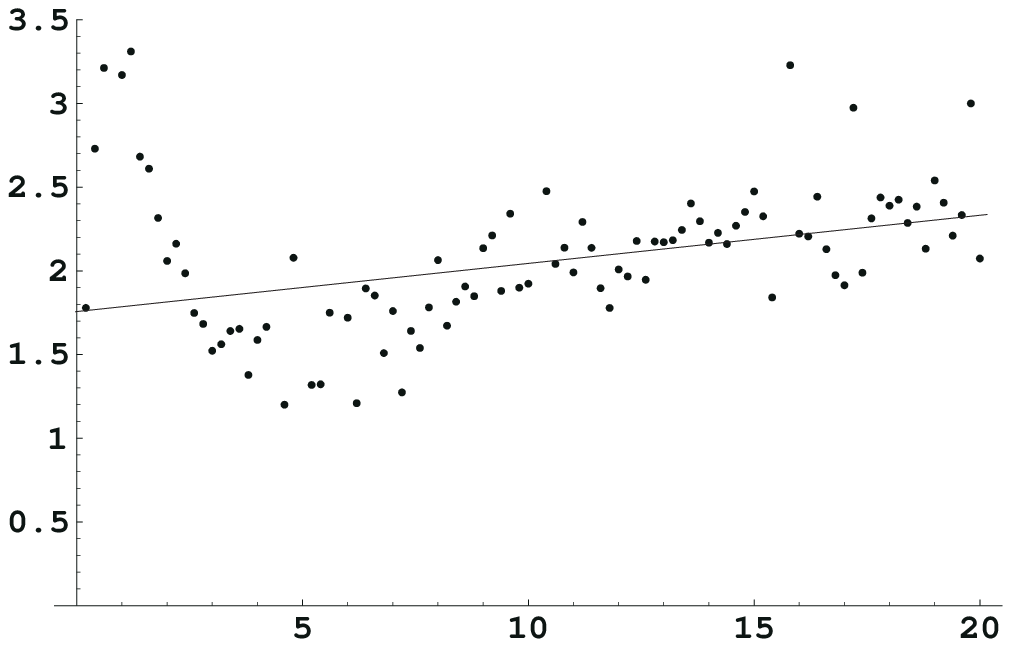}
  \caption{The ratio ${\cal E}_Q/{\cal E}_C$ as a function of $J$ for $g=1$.}
  \label{eff}
\end{figure}

Figure~\ref{eff} compares the efficiencies
${\cal E}_C$ of simulations about $\phi_\mathrm{class}$ and
${\cal E}_Q$ for simulations about $\varphi$ for various values of $J$. 
It is seen that we get a two fold improvement in efficiency. This may
not seem spectacular, and in $d=0$ it realy is not. However, once we
consider theories in $d>0$ we are dealing with true path integrals. If we
approximate the path integral with $N$ integrals then the expansion around
$\varphi$ gives a jump in efficiency of $2^N$. Even for a modest simulation
with $N=20$ this corresponds to an increase of six orders of magnitude.

The problem with this calculation is that we already need to have the exact
result for $\varphi$ in order to get the stated increase in speed.
The way out is obvious and
is reminiscent of the step we made in the previous section in going from
(\ref{w_q}) to (\ref{phi-n}). Therefore, we need to start Monte Carlo with
a Gaussian distribution centered about $\phi_\mathrm{class}$. After a while
this gives us an approximation to $\varphi$, say $\varphi^\mathrm{mc}_1$.
Using this as the center of a new probability distribution we obtain
$\varphi^\mathrm{mc}_2$, etc. Unlike the series $\varphi_0$, $\varphi_1$,
$\varphi_2,\ldots$ of the previous section, this one necessarily converges
to the exact result --- even ordinary Monte Carlo does that. The improved
Monte Carlo scheme, however, can be tailored to yield an efficiency very
near to the ideal value ${\cal E}_Q$ \cite{next}.

\section{Conclusion}

We have looked at two different ways how one can take advantage of
the (rather intuitive) fact that in quantum field theory Gaussians are best
centered about the average field $\varphi\equiv\langle \phi\rangle$.
We cast the Gaussian approximation about $\varphi$ as a recursive relation.
Working on $\phi^4$ theory in $d=0$ we have shown that the iterates of
this equation present better and better approximations to $W(J)$. This
sequence of approximations ends with $W_\infty(J)$, i.e. the best Gaussian
approximation. The first iterate in this sequence is the standard one-loop
result. The second ($W_2$) is not much more complicated, and is already
much better than the one-loop approximation. In looking at theories in
$d>0$ it may be possible to use $W_2$ to get a better analytic approximation
to (say) the effective potential.

The second use for the newly centered Gaussians is in Monte Carlo
calculations. The Monte Carlo algorithm is most efficient when
one generates random numbers through Gaussian probability distributions
centered about $\varphi$. We have shown that this can substantialy
speed up the algorithm. For an $N$-fold integral the speed up is roughly
$2^N$. We are currently working on applying the improved Monte Carlo
algorithm to models in $d\ge 1$.

\end{document}